\documentclass[conference, 10pt, letterpaper]{ieeeconf}
\usepackage{cite}
\usepackage{graphicx, epstopdf}
\usepackage{amsmath, amssymb}

\usepackage{etoolbox}
\patchcmd{\thebibliography}{\clubpenalty4000}{\clubpenalty10000}{}{}
\patchcmd{\thebibliography}{\widowpenalty4000}{\widowpenalty10000}{}{}

\title{Observability Conditions and Sensing Quality for
Unicycle Systems with Constant External Forcing 
}

\IEEEoverridecommandlockouts
\author{Natalie Brace and Kristi A. Morgansen \thanks{The authors are with the
William E. Boeing Department of Aeronautics and Astronautics at the University of Washington in
Seattle, WA, USA; email: \{nbrace, morgansn\}@uw.edu. This work has been partially supported by the National Science Foundation Graduate Research Fellowship Program under grant number DGE-1256082 and the Air Force Office of Scientific Research grant number FA9550-14-1-0398.}
}

\newcommand{\mtx}[1]{\begin{bmatrix}#1\end{bmatrix}}

\newcommand{\fb}{\mathbf{f}}
\newcommand{\hb}{\mathbf{h}}

\newcommand{\xb}{\mathbf{x}}
\newcommand{\yb}{\mathbf{y}}

\newcommand{\dO}{d\mathcal{O}}

\newcommand{\pd}[2]{\frac{\partial #1}{\partial #2}}

\newcommand{\eps}{\varepsilon}

\newtheorem{theorem}{Theorem}
\newtheorem{lemma}{Lemma}

\newtheorem{corollary}{Corollary}

\usepackage{comment}
\usepackage[usenames,dvipsnames]{xcolor}

\newcommand{\new}[1]{#1}
\newcommand{\rem}[1]{}
\newcommand{\remmath}[1]{}

\begin{document}

\maketitle

\begin{abstract}
In certain systems which are subject to significant constant external forcing
such as an airplane in
wind or an underwater glider in ocean
currents, the ability to detect the forcing
depends on both the measurements available and whether appropriate control is being applied.
Using analytical nonlinear observability techniques, we define the necessary characteristics required of a measurement function for systems with unicycle dynamics subject to constant forcing
to be observable.
We further consider the necessary associated motion characteristics required of this class of 
systems to enable the desired sensing capabilities.
We then apply these results in combination with the empirical Gramian to quantify relative observability, optimal sensor selection, and sensing quality for  different motion primitive modes 
for a Dubins path.

\end{abstract}

\section{Introduction}

The interaction of control and sensing in nonlinear systems provides interesting challenges and opportunities when considering observability, the ability to determine a system's state from available measurements. Contrary to linear systems, the structure  of nonlinear system dynamics may require or benefit from particular motion for effective sensing. 
For example, the \textit{Aerosonde} (an unmanned fixed wing vehicle) was able to identify wind speed and direction while flying across the Atlantic  using GPS, but no magnetometer, by adding curves to its path \cite{Hol01}. 
This coupling of motion with sensing is well-known in nature as the active sensing employed by animals for binaural localization, casting to track scent in a plume, and depth from monocular vision \cite{van2015mosquitoes,Floris2014}.  
In recent years, mathematical exploration of these capabilities has begun to characterize when motion can improve, or is necessary for, effective sensing \cite{Floris2014}, \cite{Hin15},  \cite{mccammon2020topology}. 

In order to expand the foundation of mathematical tools to inform the joint tasks of sensor selection and motion planning, we consider here the development of necessary conditions for observability of a class of systems in SE(2), the planar unicycle, subject to external forcing. 
This class encompasses a wide range of autonomous vehicles, such as the \textit{Aerosonde}, used in monitoring and data collection tasks, and the external forcing broadens the observability results to include consideration of typical operational constraints of fielded vehicles such as wind or ocean currents.

To investigate the observability of this class of systems, we employ two methods: nonlinear observability analysis and the empirical observability Gramian.
The former allows us to develop analytical necessary conditions for observability, and the latter allows us to consider quantitative measures of observability which facilitates sensor selection \new{and sensing quality. As shown in \cite{Pow15}, for linear systems with noise, the stochastic empirical observability Gramian provides an upper bound for the Fisher information matrix which in turn bounds the error covariance in the state estimator.  Improvements in system observability are, therefore, a key method for improving performance and robustness of state estimation.}

To analytically assess nonlinear local observability, we employ the rank condition on the observability codistribution \cite{Nij90}. 
The observability codistribution is analogous to the observability matrix in linear systems, but \rem{unlike in a linear system where observability is a global system property,} in a nonlinear system the observability is generally a local property that depends on initial conditions and applied inputs.
\new{Much previous work has concentrated on observability with range-only measurements for unicycle or tracker-target systems with unperturbed \cite{Song1999, Crasta2018} and perturbed dynamics \cite{Gadre2005}, and a study of observability for specific sensors and controls was performed for a pair of standard unicycle systems in \cite{Martinelli2005}.   The analysis  here extends the prior work with necessary conditions for general measurement function requirements.}

The empirical observability Gramian provides a means to quantify observability by directly relating perturbations in initial conditions to output energy. This method also has the advantage of avoiding calculation of a state transition matrix and has been applied to studies of both linear and nonlinear observability \cite{Kre09}. The empirical observability Gramian has been used to determine optimal control, specifically with respect to a single ranging beacon \cite{Arr15}, and to determine trajectories to prevent degenerate measurements \cite{Ron16}.

%
The practice of sensor selection and placement has been widely studied in the context of wireless sensor networks \cite{You08}, of observability analysis in the 
monitoring of traffic congestion \cite{Con16}, 
and of the empirical Gramian to identify optimal sensor types and locations on the wings of a Hawkmoth \cite{Hin15}.
Investigating observability of hawkmoth motion,
Mohren et al found neurologically inspired signals that make the system more observable with fewer sensors \cite{Moh18}, essentially 
assessing both the best sensor placement and measurement function for observability.
The ability to maintain observability despite sensor failures was studied for linear systems in \cite{Com08}; nonlinear systems have additional flexibility since their observability depends not only on sensors but can can also depend on the control/trajectory.

The specific contributions of the work in this paper are to address the characterization of necessary features of sensor selection and observability for this class of nonlinear systems and to utilize the empirical observability Gramian to implement a sensor selection strategy along Dubins paths.
The paper is organized as follows: Section II provides an overview of observability measures, both analytical and based on the empirical Gramian. 
Section III describes the system model, and Section IV presents results for nonlinear observability analysis using the matrix rank condition.
Section V investigates the system more quantitatively using the empirical observability Gramian, followed by a discussion of sensor selection and implementation in simulation in Section VI, and conclusions in Section VIII. 

\section{Nonlinear Observability} \label{sec-NLobsv}

We consider here nonlinear systems in control affine form,
\begin{equation}
     \dot{\xb} = \fb_0(\xb) + \sum_{i=1}^m \fb_i(\xb) u_i, ~
    \yb = \hb(\xb) \label{eq-controlAffine}
\end{equation}
where $\xb \in \mathbb{R}^n, u_i \in \mathbb{R}, \yb \in \mathbb{R}^{q}$, and we are interested in investigating the observability of the system with respect to control and sensor inputs. The methods used here to determine if a system is locally observable with a given sensor measurement are the rank condition of the observability Lie algebra and the empirical observability Gramian.

\subsection{Analytic Observability}
 A system is said to be observable if for every admissible input $u$, the outputs for any two states $\xb_1$ and $\xb_2$ being equal, that is $h(\xb_1, u, t) = h(\xb_2, u, t)$ for $t \geq 0$, implies that $\xb_1 = \xb_2$ \cite{Nij90}. While it may not be possible to determine global observability for a nonlinear system, the implications of two states being indistinguishable can be restricted to some neighborhood around $\xb_0$ and the local observability can be assessed using the observability Lie algebra given by
\begin{align}
  \mathcal{O} =  \{\mathbf{h}, L_{\fb_0} \mathbf{h}, L_{\fb_0}^2 \mathbf{h}, L_{\fb_0}L_{\fb_1}\mathbf{h}, \cdots \}, \label{eq-obsvLA}
\end{align}
where the components are Lie derivatives calculated as
\begin{align*}
    L_{\fb_0}\mathbf{h} = \pd{\mathbf{h}}{\xb} \fb_0, ~ 
    L_{\fb_0}^k\mathbf{h} = \pd{\mathbf{L_{\fb_0}^{k-1}}}{\xb} \fb_0, ~
    L_{f_i}L_{\fb_0}\mathbf{h} = \pd{\mathbf{L_{\fb_0}}}{\xb} f_i\dots.
\end{align*} 
The observability codistribution, $d\mathcal{O}$, is defined as the Jacobian of $\mathcal{O}$ with respect to $\xb$.
If $d\mathcal{O}$ has dimension $n$ at a given point $\xb_0$, i.e. is full rank, then the system is locally observable at $\xb_0$ \cite{Nij90}.

\subsection{Empirical observability Gramian}

To further investigate the observability of a system beyond the binary answer provided by the rank condition, the empirical observability Gramian can by employed to 
provide quantitative information regarding how observable the system is and which states are more readily observed than others.
The empirical observability Gramian, $W^{\eps} \in \mathbb{S}^n_+$, affords a method to approximate an observability Gramian for both linear and nonlinear systems based on the differences in measurements from perturbing the initial conditions, $\xb_0$, by a small amount, $\eps$, in each basis direction, $\hat{\mathbf{e}}_i$ where $\mathbf{\hat{e}}_i \in \mathbb{R}^n$ is a unit vector in the $i$th state dimension\cite{Kre09}. For a system with continuous measurements, $W^{\eps}$ is given by the integral from initial time, $t = 0$, to final time, $t_f$,
\begin{equation}
  W^{\eps}[t_f,\mathbf{x}_0] = \frac{1}{4 \epsilon^2} \int_{0}^{t_f} Y^{\eps}[t,\mathbf{x}_0,u]^T Y^{\eps}[t,\mathbf{x}_0,u] dt. \label{eq-empc} 
\end{equation}
The vector $Y^{\eps}[t,\mathbf{x},u] = \begin{bmatrix} \Delta y^{\pm 1} & \cdots & \Delta y^{\pm n} \end{bmatrix} \in \mathbb{R}^{1\times n}$ is comprised of the difference in the scalar outputs, $\Delta y^{\pm i} = y^{+i} - y^{-i}$,
%
 where $y^{\pm i}$ represents the output for trajectories generated by perturbing the initial conditions by $\pm \eps \mathbf{\hat{e}}_i$.
 
If $W^{\eps}$ is full rank, the system is locally weakly observable 
\cite{Pow15}: that is, at each point $x_0$ there exists an open neighborhood $U$ of $\xb_0$ such that for every open neighborhood $V \subset U$ of $\xb_0$, with trajectories that remain in $V$ for all $t \in [0, T]$ the output $h[t, \xb_0, u] = h[t,\xb_1,u]$ for every admissible input $u$ and pair of states $\xb_0, \xb_1 \in V$ implies that $\xb_0 = \xb_1$. \new{Here, we do not explicitly consider time or control in the measurement function and use the notation $h(\xb)$; the notation $h(x_1, \cdot)$ indicates $h$ is a function of $x_1$ and possibly other states.}
For an observable system, one is then concerned with how well it can be observed. The maximum and minimum eigenvalues of $W^{\eps}$ describe the energy in the most and least observable modes, respectively. 

\section{System model}
In this work we are specifically interested in nonlinear, control-affine systems with the addition of external forcing functions. Let the original system with drift and control inputs be defined as
\begin{align}
    \dot{\xb}_d &= \bar{\fb}_d(\xb_d) + \sum_{i=1}^m \bar{\fb}_i(\xb_d) u_i, \quad 
    y = \hb(\xb_d), \nonumber
\end{align}
$\xb_d \in \mathbb{R}^{n_d}$, $u_i \in \mathbb{R}$, $\yb \in \mathbb{R}^q$. For this paper we assume that the external states cannot be measured directly and are subject to significantly slower dynamics than the original system so they can be modeled as constant. 
Adding the additional states, $\xb_e \in \mathbb{R}^{n_e}$, the full system can be written as 
\begin{align}
    \Sigma: \quad  
    \dot{\xb} &=\mtx{\dot{\xb_d} \\ \dot{\xb_e}} = \mtx{\bar{\fb}_d(\xb_d) + A \xb_e \\ \mathbf{0}^{n_e}} + \sum_{i=1}^m \mtx{\bar{\fb}_i(\xb_d) \\ \mathbf{0}^{n_e}} u_i  \nonumber \\
   \yb &= \hb(\xb_d) \label{eq-extendedSystem}
\end{align}
 where $\xb \in \mathbb{R}^n$, $n = n_d + n_e$ and $A \in \mathbb{R}^{n_d \times n_e}$.
To simplify the notation, let 
\begin{align}
    \fb_0(\xb) &= \mtx{\bar{\fb}_d(\xb_d) + A \xb_e \\ \mathbf{0}^{n_e}}, \text{ and } \fb_i(\xb_d) = \mtx{\bar{\fb}_i(\xb_d) \\ \mathbf{0}^{n_e}}. \nonumber
\end{align}
To facilitate further discussions regarding requirements for local observability of the system \eqref{eq-extendedSystem}, the states are partitioned into the following sets: set $X:= \{X_0, X_1, ..., X_m \}$ are states that explicitly appear in the drift or control vector fields, while $\fb_0(\xb)$ or $\fb_i(\xb_d)$, (a subset of $\xb_d$ and all $\xb_e$ states) and set $Z$ are states that do not (remaining $\xb_d$ states). 

\begin{figure}
    \centering
    \vspace{0.1in}
    \includegraphics[width=0.18 \textwidth]{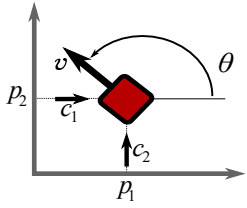}
    \vspace{-0.1in}
    \caption{Unicycle model in the $x_1-x_2$ plane with heading $\theta$ and constant forces $c_1$ and $c_2$.}
    \label{fig-unicycle}
    
    \vspace{-2em}
\end{figure}

\emph{Unicycle with external forcing:} 
As an example, consider a planar unicycle system whose location and orientation in the $x_1{-}x_2$ plane is given by $\xb_d = (p_1, p_2, \theta) \in SE(2)$ and is subject to constant but unknown forces $\bf{c} = \mtx{c_1 &c_2}^{\top}$ in the $x_1$ and $x_2$ directions.
The inputs are linear velocity, $u_1$, and angular velocity, $u_2$. The control affine system is
\begin{eqnarray}
 \Sigma_U: ~  \mtx{\dot{\xb}_d \\ \dot{\bf{c}}} = 
  \mtx{\dot{p}_1 \\ \dot{p}_2 \\ \dot{\theta} \\ \dot{c_1} \\ \dot{c_2}} &=& 
  \underbrace{\mtx{ c_1 \\ c_2 \\ 0 \\ 0 \\ 0}}_{\fb_0(\xb)} + 
  \underbrace{\mtx{ \cos (\theta)\\  \sin (\theta)\\ 0 \\ 0 \\ 0}}_{\fb_1(\xb)} u_1+ 
  \underbrace{\mtx{0 \\ 0 \\ 1 \\ 0 \\ 0}}_{\fb_2(\xb)} u_2,  \nonumber\\
  {\bf y} &=& \hb(\xb_d) \label{eq-unicycle}
\end{eqnarray}
with $\xb, \fb_0(\xb), \fb_1(\xb), \fb_2(\xb) \in \mathbb{R}^5$ and $\hb_U(\xb) \in \mathbb{R}^q$.
The measurements are restricted to functions of states in $\xb_d$. For this example system, the states are partitioned into sets $c_1, c_2 \in X_0$, $\theta \in X_1$, and $p_1,p_2 \in Z$.
 A diagram of the system is shown in Fig.~\ref{fig-unicycle}.
 
 A unicycle moving at constant speed $u_1=v$, as when following a Dubins path, includes the linear velocity in the drift:
 \begin{align}
     \Sigma_D: ~ \dot{\xb} = \big(\fb_0(\xb) + v \fb_1(\xb_d) \big)+ \fb_2 u_2 = \bar{\fb}_0(\xb) + \fb_2 u_2. \label{eq-dubins}
 \end{align}

\section{Analytical Observability Analysis} \label{sec-analysis}

Since the measurements are limited to being functions of the original $\xb_d$ states, the components of the observability Lie algebra \eqref{eq-obsvLA} can be calculated with $\fb_d(\xb)$ in place of $\fb_0(\xb)$. For example, the first Lie derivative of the $j$th measurement function, $h_j$, is calculated as
\begin{align}
    L_{\fb_0} h_j &= \pd{h_j(\xb_d)}{\xb} \fb_0(\xb) 
    = \pd{h_j(\xb_d)}{\xb_d} (\bar{\fb}_d(\xb_d) + A \xb_e)  \nonumber \\
   &= \sum_{i=1}^n f_{0i}\pd{h_j}{x_i} = L_{\bar{f}_d}h_j + \sum_{i=1}^{n_d} (A_i x_e)\pd{h_j}{x_i}. \label{eq-Lf0opNotation}
\end{align}
The second line is given in operator notation, which represents the vector field $\fb_0 = \mtx{f_{0,1} &\cdots& f_{0i} &\cdots& f_{0n} }^T$ as the sum $\fb_0 = f_{0,1} \pd{}{x_1} + \cdots +  f_{0i}\pd{}{x_i} + \cdots + f_{0n}\pd{}{x_n}$. For the extended system \eqref{eq-extendedSystem}, we have $f_{0i} = (\bar{f}_{di} + A_i \xb_e)\pd{}{x_i}$ for $i \leq n_d$ and $f_{0i} = 0$ for $i > n_d$, where $A_i \in R^{1 \times n_e}$ denotes the $i$th row of $A$. The $k$th order Lie derivative of $h_j$ with respect to $\fb_0$ is denoted $L_{\fb_0}^k h_j$.

The following lemmas define certain characteristics of the measurement function that are required for the nonlinear forced system $\Sigma$ to be locally observable. 
To provide definitive statements for general measurement functions, it is more mathematically tractable to phrase the lemmas in terms of what would prevent the system from being observable.

\begin{lemma} \label{lem-setZ}
A system $\Sigma$ will be unobservable if each state $x_i \in Z$ does not explicitly appear in $\hb(\cdot)$. 
\end{lemma}

\begin{proof}
The observability codistribution, denoted simply as $\dO$, is given by
\begin{align*}
    \dO &\equiv \pd{\mathcal{O}}{\xb}
    = \pd{}{\xb}\mtx{\hb & \pd{\hb}{\xb}\fb_0 & \cdots}^T,
\end{align*}
with terms from the observation Lie algebra calculated as in \eqref{eq-obsvLA}.
By definition, $\pd{\fb_0}{x_j} = 0$ for all $x_j \in Z$ so any column of $\dO$ corresponding to $x_j\in Z$ will be $\mathbf{0}$ if $x_j$ does not appear explicitly in $\hb$, thus ensuring rank$(\dO)<n$ and making the system unobservable.
\end{proof}

\begin{corollary} 
If a state that does not appear in the drift vector field, $x_j \in X\setminus X_0$, then it must appear in $\hb(x_j,\cdot)$ or a control corresponding to a control vector field including the state, i.e. $u_k$ for $x_j \in X_k$, must be active or the system will be unobservable.
\end{corollary}

\begin{corollary} \label{lem-x1x2}
Systems $\Sigma_U$ and $\Sigma_D$ will not be observable if their measurement function(s) do not include states $p_1$ and $p_2$. If $u_1 = 0$ and $\Sigma_U$'s measurement function(s) do not include $\theta$ it will also be unobservable.
\end{corollary}

This result is not surprising: if the dynamics are not directly influenced by a state, such as the location within the $x_1-x_2$ plane in the unicycle example, then it is not possible to observe that state through the evolution of the system's dynamics and it must be measured directly. 

\begin{lemma} \label{lem-u0onemeas}
A system $\Sigma$ with $q=1$ and $u_i = 0 ~\forall i,t$ will be unobservable if $L^{k}_{\fb_0} h(\xb_d) = 0$ for $k < n-1$.
\end{lemma}

\begin{proof}
With a single measurement function and no control vector fields, the elements of $\mathcal{O}$ are restricted to the measurement function, $h(\xb_d)$, and its subsequent Lie derivatives with respect to the drift only.
The $k$th Lie derivative, $L_{\fb_0}^k h(\xb_d)$, is composed of $k$th order derivatives of $h(\xb_d)$ with respect to $\xb$. For $\dO$ to be rank $n$ and fulfill the rank condition for local observability, the observability Lie Algebra must contain at least $n$ nonzero elements: the measurement function plus $n-1$ nonzero Lie derivatives.
\end{proof}

This lemma is the nonlinear equivalent of the requirement for a linear system, $\dot{\xb} = F \xb, y = C\xb$, that if $C$ is a single row, then $C F^{n-1} \neq 0$.

\begin{corollary} \label{cor-1meas}
System $\Sigma_U$ or $\Sigma_D$ with $q = 1$ and $u_1,u_2=0$ will be unobservable if the fourth order derivatives of $h_D(p_1, p_2,\theta)\in \mathbb{R}$ with respect to $p_1$ and/or $p_2$ is zero.
\end{corollary}


The observability framework for a general nonlinear system $\Sigma$ lacks sufficient structure to obtain further meaningful results, so we proceed by investigating the conditions for which the example systems $\Sigma_U$ and $\Sigma_D$ are locally observable through a series of lemmas.

\subsection{Unicycle system, constant speed and heading}
The next set of lemmas addresses the requirements for observability of system $\Sigma_D$ when $u_2=0$.

\begin{lemma} \label{lem-u0theta}
If $u_1\neq u_1(t)$ and $u_2 = 0$, system $\Sigma_D$ will be unobservable if $\hb$ is not a function of $\theta$.
\end{lemma}

\begin{proof}
If $\theta$ is not included in a measurement function and $u_1 = v$ is a constant,
the columns of $\dO_{D}$ are related through
\begin{align}
    \pd{\mathcal{O}_{D}}{\theta}& + \alpha_4 \pd{\mathcal{O}_{D}}{c_1} + \alpha_5 \pd{\mathcal{O}_{D}}{c_2} = 0 \label{eq-dOcols}
\end{align}
indicating that these columns are linearly dependent and $dO_{D}$ cannot have full rank. Furthermore, by calculating the partial derivatives of $\fb_0$ with respect to $\theta, c_1$ and $c_2,$ the coefficients in are $\alpha_4 = v \sin(\theta)$ and $\alpha_5 = v \cos(\theta)$. 
\end{proof}

In other words, \rem{the} Lemma \ref{lem-u0theta} demonstrates that when  heading angle and linear velocity are a constant, the contribution to the drift from $\theta$ is indistinguishable from that of the external forces $c_1$ and $c_2$, so a measurement of the heading is required for observability.
Similarly, the position states must be sufficiently differentiable that they can be individually distinguished and allow the contributions of the external forces to be measured as they propagate through the system dynamics, as further demonstrated in the following lemma.

\begin{lemma} \label{lem-u0twomeas}
System $\Sigma_D$ with $u_2 = 0$ is unobservable if $q=2$ and either of the following conditions are met:
\begin{enumerate}
    \item $h_1 = h_1(p_1,\cdot)$, $h_2 = h_2(p_2, \cdot)$, and the second derivatives of $h_i$ with respect to $p_1$ and $p_2$ are all zero
    \item $h_1 = h_1(p_1, p_2, \cdot)$, $h_2 = h_2(\theta)$, and the third derivative of $h_1$ with respect to $p_1$ and $p_2$ is zero. 
\end{enumerate}
\end{lemma}

\begin{proof}
These requirements stem from the number of Lie derivatives required for $\dO$ to be full rank; the second and third Lie derivatives are as follows:
\begin{align}
     L_{\fb_0}^2 h_i &= \pd{^2 h_i}{p_1^2}f_{0,1}^2 + 
      2\pd{^2 h_i}{p_1 p_2} f_{0,1} f_{0,2}+ \pd{^2 h_i}{p_2^2} f_{0,2}^2, \nonumber \\
     L_{\fb_0}^3 h_i &= \pd{^3 h_i}{p_1^3}f_{0,1}^3 + 
      3\pd{^3 h_i}{p_1^2 p_2} f_{0,1}^2 f_{0,2} \nonumber\\ &\qquad +  
      3\pd{^3 h_i}{p_1 p_2^2} f_{0,1} f_{0,2}^2 
      + \pd{^3 h_i}{p_2^3} f_{0,2}^3,
\end{align}
where $f_{0,i}$ is the $i$th element of $\fb_0(\xb)$.

For two measurement functions, $\mathcal{O}$ must include elements from $\hb$, $L_{\fb_0} \hb$, and $L_{\fb_0}^2 \hb$ to have a total of five components and possibly meet the rank condition for observability.
If $h_2$ is only a function of $\theta$, $L_{\fb_0} h_2(\theta) = 0$ so $L_{\fb_0}^3 h_1 \neq 0$ will be required.
If neither $p_1$ nor $p_2$ has a nonzero second (case 1) or third (case 2) derivative, then $L_{\fb_0}^2 \hb$ or $L_{\fb_0}^3 \hb$, respectively, would be zero and the maximum possible rank of $\dO$ would be four.
\end{proof}

\begin{lemma} \label{lem-u0threemeas}
System $\Sigma_D$ with $u_2 = 0$ is locally observable if $q\geq3$ and $h_1 = h_1(p_1,\cdot)$, $h_2 = h_2(p_2, \cdot)$, $h_3 = h_3(\theta,\cdot)$ are independent functions in $p_1$, $p_2$, and $\theta$. %
\end{lemma}

\begin{proof}
Due to the requirement that the elements of $\hb_D$ are independent, the first three rows of $d\mathcal{O} = \pd{\hb_D}{\xb}$ will be linearly independent. The first Lie derivatives have the form
\begin{align}
    L_{\fb_0}h_i &= \pd{h_i}{p_1}(v \cos(\theta) + c_1) + \pd{h_i}{p_2} (v \sin(\theta) + c_2). \nonumber 
\end{align}
Now consider the submatrix of $d\mathcal{O}$ that corresponds to the derivatives of the Lie derivatives $L_{\fb_0}h_1$ and $L_{\fb_0}h_2$ with respect to the external forces $c_1$ and $c_2$: 
\begin{align}
     d\mathcal{O}_{4-5,4-5} = 
     \mtx{\pd{h_1}{p_1} & \pd{h_1}{p_2} \\ 
          \pd{h_2}{p_1} & \pd{h_2}{p_2}}. \nonumber
\end{align}
Since $h_1$ and $h_2$ are independent functions with respect to $p_1$ and $p_2$, these derivatives will also be independent and this submatrix will be full rank, making $d\mathcal{O}$ full rank.
\end{proof}

Combining the previous lemmas, we can state the following theorem regarding observability of \eqref{eq-unicycle} without control:
\begin{theorem} \label{thm-linear}
System $\Sigma_D$ with $u_2 = 0$ is unobservable if either of the following conditions holds:
\begin{enumerate}
    \item $\hb_D \neq \hb_D(p_1, p_2, \theta)$
    \item $\hb_D$ does not have sufficient nonzero derivatives with respect to the states $p_1$ and $p_2$ to meet the conditions of lemmas \ref{lem-u0onemeas}, \ref{lem-u0twomeas}, and \ref{lem-u0threemeas}.
\end{enumerate}

\end{theorem}

\subsection{Unicycle system, with control}

If either control is active, the nonlinearity of the system allows it to be observable with fewer measurements, as shown in the following lemma.

\begin{lemma} \label{lem-unonzero}
Systems $\Sigma_U$ and $\Sigma_D$ can be locally observable if $u_1 = u_1(t)$ for $\Sigma_U$ and $u_2 \neq 0$ for $\Sigma_D$ and $h_1(p_1, \cdot)$ with $h_2(p_2, \cdot)$. \rem{with $L_{\fb_0}h_i \neq0$ for  $i=1,2$.}
\end{lemma}

\begin{proof}
The simplest possible measurement functions are $h_1 = p_1$ and $h_2 = p_2$. With these measurements, the observability algebras are
$\mathcal{O}_U = \{\mathbf{h}, L_{\fb_0}\mathbf{h}, L_{\fb_1} \mathbf{h}\}$ and 
$\mathcal{O}_D = \{\mathbf{h}, L_{\fb_0}\mathbf{h}, L_{\fb_0}L_{\fb_2} \mathbf{h}\}$
\begin{align}
    \dO &=  
    \mtx{1 & 0 &  & 0 & 0 \\
         0 & 1 &  & 0 & 0 \\
         0 & 0 & \pd{\mathcal{O}}{\theta} & 1 & 0 \\
         0 & 0 &  & 0 & 1 \\
         0 & 0 &  & 0 & 0 \\
         0 & 0 &  & 0 & 0} \label{eq-codist},~
        \pd{\mathcal{O}_U}{\theta} = \mtx{0 \\ 0 \\ 0 \\ 0 \\ \cos(\theta) \\ \sin(\theta)}, ~\\
        \pd{\mathcal{O}_D}{\theta} &= v\mtx{0 & 0 & -\sin(\theta)&  \cos(\theta)&  \cos(\theta)&  \sin(\theta)}^{\top}\nonumber
\end{align}
which is full rank, making the system observable.
\end{proof}

Note that if the control is not active, the last two rows of $\dO$ in \eqref{eq-codist} will be zero, rendering the system unobservable. More generally, we can combine Theorem \ref{thm-linear} with the previous lemma to state the following:

\begin{theorem}  \label{thm-withu}
System $\Sigma_U$ ($\Sigma_D$) with observation window ${\cal T}=[0,T]$ is unobservable if neither of the following conditions is met:
\begin{enumerate}
    \item $u_1 = u_1(t)$ ($u_2 \neq 0$) \rem{$u_(t)\neq 0$} for some $t \in {\cal T}$ and $\hb=\hb(p_1,p_2)$ 
    \item $\hb=\hb(p_1,p_2,\theta)$ 
\end{enumerate}
\end{theorem}

\textit{Remark:} The results of this section indicate that the unicycle system requires either sufficient measurements or nonlinearity to be observable: if the system behaves like a linear system -- as it does when $u = 0$ -- more measurements are required for observability than when it is behaving nonlinearly. 

\new{If the perturbations were not constant, additional states could be added to capture their dynamics. Correspondingly, the system would require either additional independent measurements or increased differentiability to be observable.}

\begin{figure}[tb]
    \centering
\includegraphics[width=0.65\columnwidth]{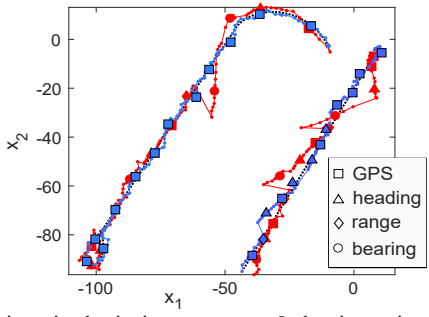}
\vspace{-0.15in}
    \caption{Trajectories beginning near $x_2 = 0$ showing estimation based on optimal (blue) and naïve (red) sensor selection. }
    \label{fig-EKFtraj}
\end{figure}
\begin{figure}[tb]
\centering
\includegraphics[width= 0.85\columnwidth]{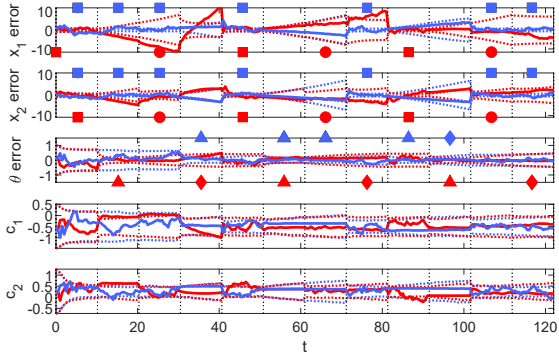}
\vspace{-0.1in}
    \caption{
Error (solid lines) and covariance (dashed lines) for the straight trajectory. The markers above/below indicate the optimal/naïve sensor selection for the segment.}
    \label{fig-EKFerror}
\vspace{-1.5em}
\end{figure}

\section{Trajectory Analysis with the Empirical Observability Gramian} \label{sec-EoG}

The results of the previous section provide base requirements that a sensor suite must meet for the system to possibly be observable, but the analytic tools do not provide a good qualitative framework for understanding how state information propagates through the dynamics to the sensors and selecting sensors based on that information. Thus we proceed by leveraging results of empirical Gramian to further explore the relation between observability and control and frame an optimization problem for sensor selection. To restrict the number of variables, in this section we consider the Dubins system.

For the empirical observability Gramian calculations in this section, the measurement functions are defined as 
\begin{align*}
    h_1 = p_1 \text{ and } h_2 = p_2.
\end{align*}
When multiple observations are used, the Gramian is the sum of the Gramians for each individual measurement \cite{Kre09}.

\subsection{Linear Trajectory}
For a linear trajectory ($u=0$, $\theta(t)=\theta_0$), the system model \eqref{eq-dubins} can be analytically integrated with respect to time.
With initial conditions $\xb_0 = \mtx{p_{1,0}&  p_{2,0}& \theta_0 & c_1 & c_2}^{\top}$,
the $Y^\eps[t,\xb_0,u]$ vectors are calculated as 
\begin{align}
    Y_{h_1}^{\eps}[t,\xb_0,0] &=\mtx{2 \eps & 0 & - 2 t v_0 \sin(\theta_0) \sin( \eps) & 2 \eps t & 0} \nonumber \\
    Y^{\eps}_{h_2}[t,\xb_0,0] &=\mtx{0 & 2 \eps & 2 t v_0 \cos(\theta_0) \sin( \eps) & 0 & 2 \eps t} \nonumber
\end{align}
and combined to form the Gramian as
\begin{align}
    W_{line} &= \frac{1}{4 \eps^2}\int_0^{t_1} (Y_{h_1}^{\eps \top} Y_{h_1}^{\eps} +  Y^{\eps \top}_{h_2} Y^{\eps}_{h_2}) dt \nonumber \\
    &= \mtx{t_1& 0& W_{1,3} & t_1^2/2&   0\\
   0&   t_1&  W_{2,3} &  0&  t_1^2/2\\
 W_{1,3} & W_{2,3}&  W_{3,3}& W_{4,3}& W_{5,3}\\
  t_1^2/2& 0& W_{4,3}&  t_1^3/3&   0\\
    0& t_1^2/2&  W_{5,3}&  0&    t_1^3/3
}, \label{eq-Wline} 
\end{align}
where
\begin{align*}
\mtx{W_{1,3} \\ W_{2,3} \\ W_{3,3} \\ W_{4,3} \\ W_{5,3} } &= \mtx{ -\frac{t_1^2 v}{2\eps} \sin(\eps)\sin(\theta_0) \\ 
\frac{t_1^2v}{2\eps}\sin(\eps)\cos(\theta_0)\\ 
\frac{t_1^3 v^2}{3\eps^2}\sin^2(\eps)(\cos^2(\theta_0) + \sin^2(\theta_0)) \\ 
-\frac{t_1^3v}{3\eps}\sin(\eps)\sin(\theta_0) \\
\frac{t_1^3v}{3\eps}\sin(\eps)\cos(\theta_0)
}.
\end{align*}
As one would expect from Theorems \ref{thm-linear} and \ref{thm-withu}, the system is not observable and this Gramian has rank $4 < n$.

\subsection{Circular Trajectory} 
For a circular trajectory with $\dot{\theta}=\omega_0$, the system model \eqref{eq-dubins} can also be integrated with respect to time.
With initial conditions $x_0 = \mtx{p_{1,0}&  p_{2,0}& \theta_0 & c_1 & c_2}^{\top}$ and constant control $u = \omega_0$, \new{
 the $Y^{\eps}[t,\xb_0,u]$ vectors are
\begin{align} 
    &Y^{\eps}_{h_1} = \mtx{2 \eps & 0 & \frac{2 v_0 \sin( \eps)\big( \cos(\omega_0 t + \theta_0) - \cos(\theta_0) \big) }{\omega_0}  & 2 \eps t & 0} \nonumber \\
    &Y^{\eps}_{h_2} = \mtx{0 & 2 \eps & \frac{2 v_0\sin( \eps)\big(\sin(\omega_0 t + \theta_0) - \sin(\theta_0) \big)}{\omega_0}  & 0 & 2 \eps t} \nonumber
\end{align}}
and combine to form 
the empirical observability Gramian, which has a similar form to \eqref{eq-Wline} with additional trigonometric terms.

\subsection{Combined paths}

\begin{figure}[tb]
   \centering
    \includegraphics[width=0.65\columnwidth]{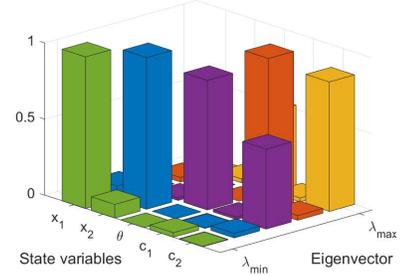}
    \vspace{-0.18in}
    \caption{Normalized eigenvectors corresponding equal time in linear and circular sections.}
    \label{fig-evecs}
\end{figure}

The results of the previous sections can be combined to calculate the Gramian for a Dubins path including circular and linear segments
\begin{align*}
    W_{Dub} = W_{circle} + W_{line}
\end{align*}
where the time spent in the circular and linear sections are given by $t_2$ and $t_1$, respectively. 
To see what information is most readily passed on through the measurements, we can look at the eigenvectors of the system. The normalized eigenvectors corresponding to equal time in each section \rem{the eigenvalues at the midpoint of Fig. \ref{fig-evalsBalance} }(when $t_1 = t_2$) are shown in Fig. \ref{fig-evecs}, which indicate that perturbations in $c_2$, $c_1$, and $\theta$ transmit more energy through the system dynamics to the outputs than do the states  being measured directly.

\section{Sensor Selection}

Section \ref{sec-analysis} provides a qualitative view of what type of measurements are required for \eqref{eq-unicycle} to be observable. In this section we will discuss what that means in terms of actual sensor selection. Some typical sensors and their measurement functions are listed in Table \ref{tab-sensors}\new{; the empirical Gramian is developed for all sensors similarly to those presented in \ref{sec-EoG}}.

For autonomous systems it is generally advantageous to minimize energy usage, which can be done by using only the necessary sensors. 
While we can mathematically describe a measurement function that uses only one ``measurement" to make the system fully observable, that is not typically feasible without combining data from multiple sensors.
Using two sensors, $\Sigma_D$ can be observable without control: the GPS (or other positioning system) with the magnetometer or bearing to beacon. If the heading sensor should fail, the system could regain observability by executing a control maneuver.

\subsection{Optimization}
To assess which sensors are best at a given time, we break a trajectory up into $K$ segments and define sensor activation variables $s$ that serve to turn on and off sensors. Let the empirical observability Gramian for sensor $i$ over the $k$th time step be given as $W_i[k]$, then the full Gramian is defined as $W(s) = \sum_{k=1}^K W_i[k] s_i[k]$ and the optimization problem is given by 
\begin{eqnarray}
 \max_{\mathbf{s}}  \lambda_{min}(W(s)) & \text{s.t.}
&\sum_{i=1}^p s_i[k] \leq 1 \quad \forall k  \\
& &  s_i[k] \in \{0,1\} \quad \forall i,k.  \nonumber 
\end{eqnarray}  
where $p$ is the number of potential sensors. 
Relaxing the requirement on $s$ to $s \in (0,1)$ the problem becomes convex.

\new{We choose $\lambda_{min}$ as the objective function since it corresponds to the smallest eigenvalue of the mapping between $\yb$ to $\xb$ and provides a measure for how easily the initial state can be recreated from the measurements \cite{Kre09}.}

\begin{table}
\vspace{0.1in}

\centering
\begin{tabular}{|l|l|c|}
\hline
\textbf{Type of Sensor} & \textbf{Measurement Function(s)}      & \new{Error $\sigma$} \\\hline
GPS              & $h_1 = p_1$ and $h_2 = p_2$  &  \new{2 m}\\ \hline
Magnetometer     & $h_3 = \theta$       & \new{$12^{\circ}$}\\  \hline
Range to beacon  & $h_4 = \sqrt{(p_1 - b_1)^2 + (p_2-b_2)^2}$ & \new{1 m} \\ \hline
Bearing to beacon& $h_5 = \text{atan2}(p_2-b_2,p_1-b_1)-\theta$  & \new{$5^{\circ}$}\\ \hline
\end{tabular}
\caption{Examples of sensors and measurement functions.\vspace{-3em}} 
\label{tab-sensors}
\end{table}

\subsection{Simulation}
The sensor selection was performed on Dubins trajectories with the sensors from Table \ref{tab-sensors}. The selected sensor data was provided to an extended Kalman Filter (EKF) using data simulated in \textsc{Matlab} (Figs. \ref{fig-EKFtraj}-\ref{fig-EKFerror}). 
To maximize the minimum eigenvalue over trajectories that include curved sections, sensors were selected to provide information on the location in the $x_1-x_2$ plane. For the straight line trajectory (bottom center), however, heading measurements were also required. In both cases, the error and generally the error covariance were better for the optimized sensor selections than for the naïve.

\section{Conclusions}
The work presented in this paper gives necessary conditions for measurement functions to make a unicycle system with constant external forcing locally observable with and without control inputs based on nonlinear analysis techniques. 
The empirical observability Gramian was then used to quantitatively study the explicit coupling between motion planning, sensor selection, and observability.

This work is an initial step towards developing a comprehensive approach to nonlinear observability and associated estimation error covariance for this class of externally forced systems that encompasses a wide variety of autonomous vehicles.
We will build upon it by continuing to refine methods for determining the coupling between sensing and motion requirements, both analytically and empirically, with an aim to develop a constructive strategy for combined sensor selection and motion planning.

\bibliographystyle{IEEEtran}
\bibliography{IEEEabrv,refs}


\end{document}